# Transport Layer Networking:

Yatish Kumar - Energy and Science Network ( ESnet )  yak@es.net ,
Stacey Sheldon - ( ESnet ) stac@es.net , Dale Carder - ( ESnet ) dwcarder@es.net

## Introduction :

In recent years, the networking industry has discovered and implemented software defined networking.  This began with OpenFlow, which was then replaced by P4, a more general programming language that can parse and execute von Neumann style processing on the packet headers and data.  Coupled with this advance, industry has also produced smartNICs ( Smart Network Interface Cards ), which are built on either FPGA ( Field Programmable Gate Array ) or NPU ( Network Processing Unit ) asics.  This provides an affordable combination of network speed and processing hardware.  In this paper we describe the opportunities that are now available for network protocol development which were lacking in the recent past due to closed vendor implementations of routers and new networking protocols.  Bespoke protocols for scientific large data networking are now possible, harkening back to the 80's when the invention of the internet protocols came from Research and Education networks.

In this paper we focus on the invention of new network forwarding behaviors between network Layers 4 and Layer 7 in the OSI network model.  Our design goal is to propose **no changes to L3** - The IP network layer, thus maintaining 100% compatibility with the existing internet.  Small changes are made to L4 the transport layer, and a new design for a  session ( L5 ) is proposed.  This new capability is intended to have **minimal or no impact on the application layer**, except for exposing the ability for L7 to select this new mode of data transfer or not.

| # | Layer | Description |
|---|---|---|
| 7 | Application Layer | Human-computer interaction layer, where applications can access the network services |
| 6 | Presentation Layer | Ensures that data is in a usable format and is where data encryption occurs |
| 5 | Session Layer | Maintains connections and is responsible for controlling ports and sessions |
| 4 | Transport Layer | Transmits data using transmission protocols including TCP and UDP |
| 3 | Network Layer | Decides which physical path the data will take |
| 2 | Data Link Layer | Defines the format of data on the network |
| 1 | Physical Layer | Transmits raw bit stream over the physical medium |

The invention of new networking technologies is frequently done in an academic setting, however the design needs to be constrained by practical considerations for cost, operational feasibility, robustness and scale.  Our goal is to improve the production data infrastructure for HEP 24/7 on a global scale, and networking protocols are simply a means to this end.

Without limiting ourselves, we describe three forwarding use cases and goals in this paper.  Others are possible leveraging the infrastructure we propose.

# HEP determined multi-domain forwarding paths

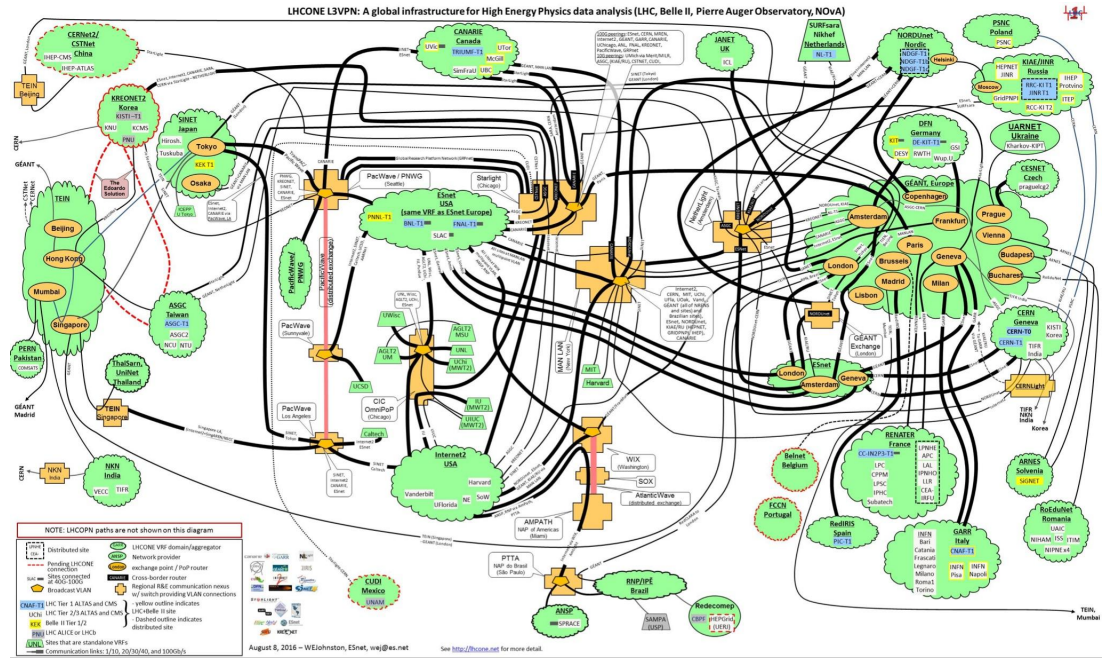

The diagram above illustrates the complexity of the network which ties together LHC data transfers.  Each green circle represents a different organization, implementing their own local decisions about how the network traffic travels within their opaque green domain, whilst ensuring that end to end signaling and data flow conform to Layer 3 Internet protocol standards.

End to end path selection between two hosts, separated by this network, is done using a set of local constraints and preferences.  The goal of the local preferences is to optimize connections with their nearest neighbors, leading to local minima in terms of the overall goal of finding not just one best path between two hosts.   The current routing implementation commonly selects a single best path, rather than using many paths between the two network domains.  If we look at link capacity, most links are operating at 50% or less, and traffic follows a single redundant path out of 2 or more backup paths.   0.5 x 0.5 = 0.25, or just ¼ of the potential throughput, and in reality much less than a ¼.  By converting to a fair, water filling session layer it is possible to recover large portions of the unused bandwidth, multiplying the value of a Billion dollar network infrastructure.

# Publish / Subscribe ( Pub/Sub ) data distribution model

Present data transfers are orchestrated between a source and a destination.   HEP has constructed a tiered ( Tier 1 , Tier 2 , Tier 3 ) data distribution architecture and the associated control plane to manage millions of jobs per hour, to fetch data from the nearest Tier of the storage hierarchy.   In recent years, the model for data lakes and streaming partial data sets has been shown to be the promising new direction for HEP storage.  Within this model, it is possible to separate the source of a dataset from the destination, making it possible for one or more destinations to subscribe to a transfer source based on network and data availability.  Some data is more popular than others, based on the science it represents.  This can be leveraged for minimizing the number of times it crosses valuable network links, such as transatlantic cables and transcontinental backbones.  Pub/Sub L5 sessions can be seen as analogous to multicast or broadcast sessions for common data between sources and sinks.

## Network Load Balancing and ESnet Network Edge Gateways

The previous two use cases describe the benefits of a coordinated global network and introducing new session layer techniques in many different locations.  In this third and final section we describe what ESnet is considering for its own network solutions.  This is an area where we can speak with more authority, since the technical decision making is local to the ESnet network domain and operating group.

ESnet is exploring the benefits of introducing "data aware" network gateways at each of our large scientific sites.  These gateways would at a minimum allow us to optimize the internal use of our own network topology through bridging a L3 customer/provider relationship between end sites and the network.  By data aware, we mean that the gateways have the attributes of a Data Transfer Node as a service.  Whilst they do not become aware of the full metadata and schema for the entire HEP data set, they do maintain an ephemeral awareness of files or data objects that are staged for data transfers across the network.

The same gateways can serve as an incubator for multi domain concepts described above.  For the snowmass planning process we advocate networking research exercises based on the assumption of having at least 1 large network provider (ESnet) provide such capabilities, with the optimistic assumption that other similarly large providers will do the same.

# Detailed Discussion and Inspirational Examples:

## A cunning plan

In the introduction we mentioned that any new network technology needs to be grounded in practical considerations for cost, respectfulness of operator administrative boundaries, transparent to the layers below, and above, and functionally compatible with the existing IP internet and applications built on it.  In order to achieve these goals, we propose the introduction of a new network element, described as a Layer 5 session layer anchor point.  This is the Layer 5 equivalent of a Layer 3 router, and is illustrated below.

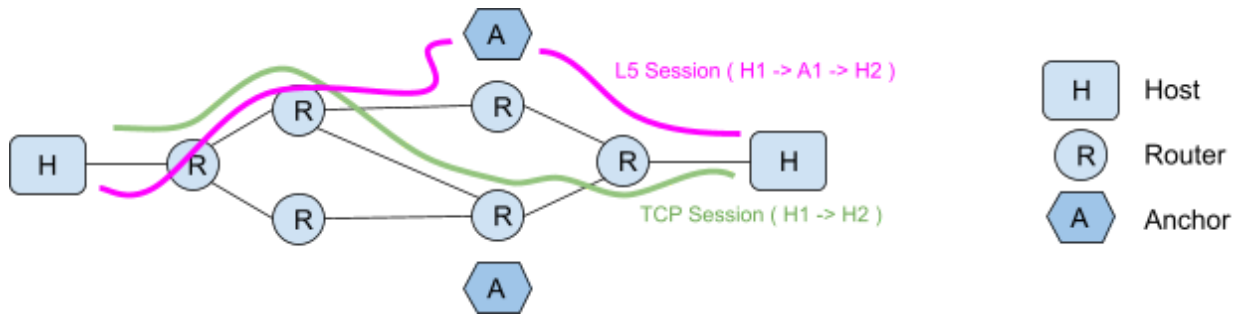

Layer 5 anchor points have the following attributes:
1. An anchor point terminates Layer 3 at its ports
2. An anchor point has a Layer 5 address
3. Hosts are visible in the Layer 5 topology and also have Layer 5 addresses
4. Anchor points use Layer 3 port connections to reach other anchor points.
5. Layer 3 routers use Layer 2 port connections to reach other L3 routers

Consider the example where two hosts, one located at CalTech on the US west coast and another located at CERN need to communicate. Presently they would exchange Layer 3 addresses, open a TCP/IP session, and the forwarding path is determined by all the intermediate networks providing the single preferred path between the 2 hosts.  The host and applications have no influence on the path selection.

With an anchor point aware L5 protocol, it is possible for the host to specify a *Layer 5 destination address* for the CERN host, but send the packet to a particular northern anchor point A.  The destination Layer 3 IP address when the packet leaves the originating host at CalTech is that of anchor point A vs the final IP address of receiving host H at CERN.  In this situation, the IP network has no choice but to send the packet to the selected anchor point, resulting in path selection at layer 5. But layer 5 equipment belongs to a HEP administrative domain, preserving the detente that exists between network operators and their customers.

In order to be feasible, these anchor points need to be affordable and compact so that they can be hosted in internet exchange points where networks meet.  They need to be strategically placed where they can influence network path selection without requiring an anchor point for every router in the network.  Finally they need to be capable, providing line rate L5 protocol forwarding with 400G and emerging 800G networking ports.

This cunning plan forms the basis for all the services detailed above.  In some protocol combinations, for example using IPv6 segment routing, it might be possible to virtualize the anchor point behavior, making them a feature in a capable router, and eliminating extra hardware.  However we will leave that for exploration in another paper.  For this paper, let us assume a physical device consisting of a 4U server with multiple 400G smartNIC cards, and very high speed packet buffers.

# Conclusion

In order to achieve the technical goals for a new session layer, it can lead to a slippery slope where L7 applications become more difficult to implement.  For example, prior efforts have led to the clear feedback that tools like Rucio and PhEDEx should not have to be topology aware, or directly involved in the control plane for data movement.  The proposed L5 session layer needs to auto discover its own topology and make its own control plane and forwarding decisions.

Managing topology discovery and doing path computation for L3 packets is the role of L3 protocols like BGP.  It is important that the L5 topology and path finding not become as complex as a full internet scale BGP implementation, requiring an additional set of new multi-domain BGP-equivalent peerings for a new network layer.

Reliable delivery and network congestion protection have been solved when using TCP as a session layer protocol.  As we re-invent the session layer concept, in order to manage multi-path, multi-domain forwarding, we need to ensure that the replacement is just as robust.

While we do not aim to burden L7 applications with L5 operational behavior, for advanced L7 use cases the following attributes can be leveraged to go beyond our present systems:

L5 end points can be identified not just by their public IP addresses, but also using a new name space.  This creates a pathway to incorporating concepts from named data networking, where both the data, as well as the end points have an identity encoded in their names.  Once we have the concept of named data, the goals of the packet marking workgroup can be further advanced, as network statistics can automatically track the science groups and their related data.

If the network behaves like a broadcast or pub/sub pipeline, it opens the door to writing applications that implement data caching and staging of data within the network.  This can support the concept of data lakes and the efficiency of demand driven replication of data elements.

The HEP community is already working towards packet marking, which allows the network to become aware of which science domain a packet of data belongs to.  This information can be recovered by a Layer 5 protocol, and using a weighted, policy driven control plane, it can assign network capacity between different science domains.  All without any change or intervention by the underlying network service provider.  This can lead to an ultimate evolution of LHCONE, where it is either not needed, or is shared by multiple science use cases.

The vision of an L5 session layer, optimized for HEP workflows, can be as simple as implementing some overlay tunnels using existing network protocols, or as complex as a complete data aware distributed storage system.  Finding the best practical implementation requires new network research and experimentation over the next 10 years.

Web References:

TCP Congestion Control : https://en.wikipedia.org/wiki/TCP_congestion_control
BGP L3 Routing Protocol: https://en.wikipedia.org/wiki/Border_Gateway_Protocol
OSI Network Model: https://en.wikipedia.org/wiki/OSI_model

Paper References: